\documentclass[12pt]{article}
\begin{document}
\title{\bf Teleparallel Energy-Momentum Distribution
of Static Axially Symmetric Spacetimes}

\author{M. Sharif \thanks{msharif@math.pu.edu.pk} and M. Jamil Amir
\thanks{mjamil.dgk@gmail.com}\\
Department of Mathematics, University of the Punjab,\\
Quaid-e-Azam Campus, Lahore-54590, Pakistan.}

\date{}

\maketitle

\begin{abstract}
This paper is devoted to discuss the energy-momentum for static
axially symmetric spacetimes in the framework of teleparallel theory
of gravity. For this purpose, we use the teleparallel versions of
Einstein, Landau-Lifshitz, Bergmann and M$\ddot{o}$ller
prescriptions. A comparison of the results shows that the energy
density is different but the momentum turns out to be constant in
each prescription. This is exactly similar to the results available
in literature using the framework of General Relativity. It is
mentioned here that M$\ddot{o}$ller energy-momentum distribution is
independent of the coupling constant $\lambda$. Finally, we
calculate energy-momentum distribution for the Curzon metric, a
special case of the above mentioned spacetime.
\end{abstract}


\section{Introduction}

Among all available theories of gravitation in the literature,
General relativity (GR) has been accepted as a true theory of
gravitation as many physical aspects of nature have been
experimentally verified in this theory. However, the localization
of energy and momentum [1] in GR is still an open, unresolved and
disputed problem. In GR, many attempts have been made to solve
this problem but no definition has generally been accepted till
now. As a pioneer, Einstein [2] used the notion of energy-momentum
complex to solve this problem. Following Einstein, many scientists
like Landau-Lifshitz [3], Papapetrou [4], Bergmann [5], Tolman
[6], Weinberg [7] and M$\ddot{o}$ller [8] have introduced their
own energy-momentum complex. All these prescriptions, except
M$\ddot{o}$ller's, are restricted to do calculations in Cartesian
coordinates only. But this difficulty was removed in
M$\ddot{o}$ller's prescription. Also, we can not define angular
momentum with the help of all these prescriptions. Misner et al.
[1] showed that energy can only be localized in spherical systems.
But later on, Cooperstock and Sarracino [9] proved that if energy
is localizable for spherical systems, then it can be localized in
any system. Bondi [10] argued that a non-localizable form of
energy is not allowed in GR.

After this, the idea of quasi-local energy was introduced by
Penrose and other scientists [11-14]. In this method , one can use
any coordinate system while finding the quasi-local masses to
obtain the energy-momentum of a curved spacetime. Bergqvist [15]
considered seven different definitions of quasi-local masses and
showed that no two of these definitions gave the same result.
Chang at el. [16] proved that every energy-momentum complex can be
associated with a particular Hamiltonian boundary term. Thus the
energy-momentum complexes may also be considered as quasi-local.
Xulu [17-19] extended this investigation and found the same energy
distribution in the case of Melvin magnetic and Bianchi type I
universe.

Virbhadra and his collaborators [20-23] verified for
asymptotically flat spacetimes that different energy-momentum
complexes can give the same result for a given spacetime. They
also found encouraging results for the case of asymptotically
non-flat spacetimes by using different energy-momentum complexes.
Aguirregabiria et. al. [24], by using the Einstein, Landau
Lifshitz, Papapetrou, Bergmann, and Weinberg (ELLPBW)
prescriptions, showed that the energy distribution within a
Kerr-Schild metric is same. Recently Virbhadra [25] found that
these five different prescriptions (ELLPBW) did not give the same
results for the most general non-static spherically symmetric
spacetime. One of the authors [26-28] found several examples which
do not provide the same result for different prescriptions. The
results [19,21,23,25-29] lead to know that the energy distribution
in M$\ddot{o}$ller's prescription is different from Einstein's
energy for some particular spacetimes, including Schwarzschild
spacetime.

Some authors [30-36] argued that this problem of energy may be
settled in the context of teleparallel theory (TPT) of gravity. They
showed that energy-momentum can also be localized in the framework
of this theory. It has been shown that the results of the two
theories agree with each other. Vargas [32] found that the total
energy of the closed Friedmann-Robertson-Walker spacetime is zero by
using teleparallel version of Einstein and Landau-Lifshitz
complexes. This agrees with the result obtained by Rosen [37] in GR.
Salti and his co-workers [33-36] considered some particular
spacetimes and calculated energy-momentum densities by using
different prescriptions both in GR and TPT and found the similar
results. Recently, Sharif and Amir [38] evaluated the
energy-momentum distribution of Lewis-Papapetrou spacetimes by using
the TP version of M$\ddot{o}$ller's prescription and found that the
results do not agree with those available in the context of GR [39].
In this paper, we investigate energy-momentum distribution for the
Weyl metrics in the context of the TPT. Further, it has been
extended to the special case as a Curzon metric. We also compare our
results with [40-42] in the context of GR.

The scheme of this paper is as follows. In section \textbf{2}, we
give some basics of TPT and TP version of Einstein,
Landau-Lifshitz, Bergmann and M$\ddot{o}$ller prescriptions.
Section \textbf{3} is devoted to the evaluation of the
energy-momentum density components for static axially symmetric
spacetimes and also for the Curzon metric. In the last section, we
shall summarize the results.

\section{TP Version of Energy-Momentum Complexes}

Before giving the TP version of the energy-momentum complexes, we
briefly outline the main points of the TP theory. The basic entity
of the theory of teleparallel gravity (TPG) is the non-trivial
tetrad [43] ${h^a}_\mu$ whose inverse is denoted by ${h_a}^\nu$.
They satisfy the following relations
\begin{eqnarray}
{h^a}_\mu{h_a}^\nu={\delta_\mu}^\nu; \quad\
{h^a}_\mu{h_b}^\mu={\delta^a}_b.
\end{eqnarray}
The theory of TPG is described by the Weitzenb$\ddot{o}$ck
connection given as
\begin{eqnarray}
{\Gamma^\theta}_{\mu\nu}={{h_a}^\theta}\partial_\nu{h^a}_\mu
\end{eqnarray}
which is obtained due to the condition of absolute parallelism
[44]. This implies that the spacetime structure underlying a
translational gauge theory is naturally endowed with a
teleparallel structure [44,45]. In this paper, the Latin alphabet
$(a,b,c,...=0,1,2,3)$ will be used to denote the tangent space
indices and the Greek alphabet $(\mu,\nu,\rho,...=0,1,2,3)$ to
denote the spacetime indices. The Riemannian metric in TPT arises
as a by product [44] of the tetrad field given by
\begin{equation}
g_{\mu\nu}=\eta_{ab}{h^a}_\mu{h^b}_\nu,
\end{equation}
where $\eta_{ab}$ is the Minkowski spacetime such that
$\eta_{ab}=diag(+1,-1,-1,-1)$. In TPT, the gravitation is
attributed to torsion [45] which plays the role of force here. For
the Weitzenb$\ddot{o}$ck spacetime, the torsion is defined as [46]
\begin{equation}
{T^\theta}_{\mu\nu}={\Gamma^\theta}_{\nu\mu}-{\Gamma^\theta}_{\mu\nu}
\end{equation}
which is antisymmetric in nature. Due to the requirement of
absolute parallelism, the curvature of the Weitzenb$\ddot{o}$ck
connection vanishes identically [43]. The Weitzenb$\ddot{o}$ck
connection and the Christoffel symbols satisfy the following
relation
\begin{equation}
{{\Gamma^{0}}^\theta}_{\mu\nu}={\Gamma^\theta}_{\mu\nu}
-{K^{\theta}}_{\mu\nu},
\end{equation}
where ${{\Gamma^{0}}^\theta}_{\mu\nu} $ are the Christoffel
symbols and ${K^{\theta}}_{\mu\nu}$ denotes the {\bf contorsion
tensor} and is given by
\begin{equation}
{K^\theta}_{\mu\nu}=\frac{1}{2}[{{T_\mu}^\theta}_\nu+{{T_\nu}^
\theta}_\mu-{T^\theta}_{\mu\nu}].
\end{equation}
The teleparallel version of the Einstein, Landau-Lifshitz and
Bergmann energy-momentum complexes, by setting $c=1=G$, are
respectively given by [32]
\begin{eqnarray}
hE_\nu^\mu&=&\frac{1}{4\pi}\partial_\lambda({U_\nu}^{\mu\lambda}),
 \nonumber\\
hL^{\mu\nu}&=&\frac{1}{4\pi}\partial_\lambda(hg^{\mu\beta}{U_\beta}
^{\nu\lambda}), \nonumber\\
hB^{\mu\nu}&=&\frac{1}{4\pi}\partial_\lambda({g^{\mu\beta}U_\beta}
^{\nu\lambda}),
\end{eqnarray}
where ${U_\nu}^{\mu\lambda}$ is the  Freud's superpotential given
as
\begin{equation}
{U_\nu}^{\mu\lambda}=h{S_\nu}^{\mu\lambda}.
\end{equation}
Here $S^{\nu\mu\lambda}$ is a  tensor quantity which is skew
symmetric in its last two indices and is defined as
\begin{equation}
S^{\nu\mu\lambda}=m_1T^{\nu\mu\lambda}+\frac{m_2}{2}(T^{\mu\nu\lambda}-
T^{\lambda\nu\mu})+\frac{m_3}{2}(g^{\nu\lambda}{T^{\beta\mu}}_
\beta-g^{\mu\nu} {T^{\beta\lambda}}_\beta),
\end{equation}
where $m_1,~m_2$ and $m_3$ are three dimensionless coupling
constants of TPG [35]. It is mentioned here that $hE_0^0$,
$hL^{00},~hB^{00}$ are the energy densities,
$hE_i^0,~hL^{0i},~hB^{0i}~(i=1, 2, 3)$ are the momentum densities
and $hE_0^i,~hL^{i0},~hB^{i0}$ are the current energy densities of
Einstein, Landau-Lifshitz and Bergmann prescriptions respectively.
Telepararllel equivalent of GR may be obtained by considering the
following particular choice [44]
\begin{equation}
m_1=\frac{1}{4}, \quad m_2=\frac{1}{2},\quad m_3=-1.
\end{equation}
The superpotential of the M$\ddot{o}$ller tetrad theory is given
by Mikhail et al. [30] as
\begin{equation}
{U_\mu}^{\nu\beta}=\frac{\sqrt{-g}}{2\kappa}P_{\chi\rho\sigma}^{\tau\nu\beta}
[{\Phi^\rho}g^{\sigma\chi} g_{\mu\tau}-\lambda g_{\tau\mu}
K^{\chi\rho\sigma}-(1-2\lambda)g_{\tau\mu} K^{\sigma\rho\chi}],
\end{equation}
where
\begin{equation}
P_{\chi\rho\sigma}^{\tau\nu\beta}= {\delta_\chi}^{\tau}
g_{\rho\sigma}^{\nu\beta}+{\delta_\rho}^{\tau}
g_{\sigma\chi}^{\nu\beta}-{\delta_\sigma}^{\tau}
g_{\chi\rho}^{\nu\beta},
\end{equation}
while $ g_{\rho\sigma}^{\nu\beta}$ is a tensor quantity and is
defined by
\begin{equation}
g_{\rho\sigma}^{\nu\beta}={\delta_\rho}^{\nu}{\delta_\sigma}^{\beta}-
{\delta_\sigma}^{\nu}{\delta_\rho}^{\beta}.
\end{equation}
$K^{\sigma\rho\chi}$ is contortion tensor as given by Eq.(6), $g$
is the determinant of the metric tensor $g_{\mu\nu}$, $\lambda$ is
the free dimensionless coupling constant of TPG, $\kappa$ is the
Einstein constant and $\Phi_\mu$ is the basic vector field given
by
\begin{equation}
\Phi_\mu={T^\nu}_{\nu\mu}.
\end{equation}
Now we can write the M$\ddot{o}$ller energy, momentum and energy
current densities as follows
\begin{equation}
\Xi_\mu^\nu= U_\mu^{\nu\rho},_\rho,
\end{equation}
where comma means ordinary differentiation. Here
$\Xi_0^0,~\Xi_i^0$ and $\Xi_0^i$ are the energy, momentum and
energy current densities respectively in M$\ddot{o}$ller's
prescription.

\section{Static Axially Symmetric Spacetimes}

The Weyl metrics are a subclass of stationary axially symmetric
spacetimes. These metrics can be reduced from the Lewis-Papapetrou
metric [47] (a class of stationary axially symmetric spacetimes)
by vanishing the angular velocity. In GR, Weyl exterior solutions
to the Einstein field equations represent all possible static
axially symmetric spacetimes [48]. They may be represented as
series expansions of suitable defined relativistic multipole
moments [49]. Thus each Weyl metric is characterized by a specific
combination of such multipoles. It would be interesting to
investigate energy-momentum distribution for this class of
spacetimes. In cylindrical coordinates $(\rho,\phi,z)$, it is
given by
\begin{equation}
ds^2=e^{2\psi}dt^2-e^{2(\gamma-\psi)}(d\rho^2+dz^2)-\rho^2e^{-2\psi}d\phi^2,
\end{equation}
where $\gamma$ and $\psi$ are functions of $\rho$ and $z$ only.
The metric functions satisfy the following constraint equations
\begin{eqnarray}
\psi_{\rho\rho} + \frac{1}{\rho}\psi_\rho + \psi_{zz}=0,\\
\gamma_\rho=\rho({\psi_\rho}^2 - {\psi_z}^2), \quad\quad
\gamma_z=2\rho\psi_\rho\psi_z.
\end{eqnarray}
Eq.(17) implies that the function $\psi$ satisfies Laplace
equation. The general solution of this equation yields an
asymptotic behavior and is given by
\begin{equation}
\psi=\sum \frac{a_n}{r^{n+1}}P_n(cos\theta),
\end{equation}
where $r=\sqrt{\rho^2+z^2}$, $cos\theta=z/r$ are Weyl spherical
coordinates and $P_n(cos\theta)$ are Lagendre polynomials. The
coefficients $a_n$ are arbitrary constants which are called
\textit{Weyl moment}.

\subsection{Energy-Momentum Densities of Einstein,
Landau-Lifshitz and Bergmann Prescriptions}

Since these prescriptions can give meaningful results only in
Cartesian coordinates thus we need to write tetrad in terms of
Cartesian coordinates. This can be obtained by writing Eq.(16) as
\begin{eqnarray}
ds^2=e^{2\psi}dt^2-\frac{1}{\rho^2}(x^2e^{2(\gamma-\psi)}+y^2e^{-2\psi})dx^2-
\frac{2xy}{\rho^2}(e^{2(\gamma-\psi)}-e^{-2\psi})dxdy\nonumber\\
-\frac{1}{\rho^2}(y^2e^{2(\gamma-\psi)}+x^2e^{-2\psi})dy^2
-e^{2(\gamma-\psi)}dz^2,
\end{eqnarray}
where $\rho=\sqrt{x^2+y^2}$. The corresponding tetrad can be
written as
\begin{equation}
{h^a}_\mu=\left\lbrack\matrix { e^\psi   &&&   0    &&&   0    &&&
0 \cr 0       &&& \frac{x}{\rho}e^{\gamma-\psi} &&&
\frac{y}{\rho}e^{\gamma-\psi}   &&&   0 \cr 0 &&&
-\frac{y}{\rho}e^{-\psi} &&& \frac{x}{\rho}e^{-\psi} &&& 0 \cr 0
&&& 0    &&& 0 &&& e^{\gamma-\psi} \cr } \right\rbrack
\end{equation}
and its inverse becomes
\begin{equation}
{h_a}^\mu=\left\lbrack\matrix { e^{-\psi}   &&   0    &&   0    &&
0 \cr 0        && \frac{x}{\rho}e^{\psi-\gamma} &&
\frac{y}{\rho}e^{\psi-\gamma}   &&   0 \cr 0 &&
-\frac{y}{\rho}e^\psi && \frac{x}{\rho}e^\psi && 0 \cr 0 &&   0 &&
0    && e^{\psi-\gamma} \cr } \right\rbrack.
\end{equation}
Here $h=det \{{h^a}_\mu\}=\sqrt{-g}= e^{2\gamma-2\psi}$. Using
Eqs.(21) and (22) in Eq.(2), we get the following non-zero
components of the Weitzenb$\ddot{o}$ck connection
\begin{eqnarray}
{\Gamma^0}_{01}&=&\frac{x}{\rho}\psi_\rho,\quad
{\Gamma^0}_{02}=\frac{y}{\rho}\psi_\rho, \quad
{\Gamma^0}_{03}=\psi_z, \nonumber\\
{\Gamma^1}_{11}&=&\frac{x}{\rho^3}(x^2\gamma_\rho-\rho^2\psi_\rho),
\quad {\Gamma^1}_{12}=\frac{y}{\rho^3}(x^2\gamma_\rho-\rho^2\psi_\rho),\nonumber\\
{\Gamma^1}_{13}&=& \frac{1}{\rho^2}(x^2\gamma_z-\rho^2\psi_z),
\quad {\Gamma^1}_{21}=\frac{y}{\rho^3} (x^2\gamma_\rho-\rho),\nonumber\\
{\Gamma^1}_{22}&=&\frac{x}{\rho^3}(y^2\gamma_\rho+\rho), \quad
{\Gamma^1}_{23}={\Gamma^2}_{13}=
\frac{xy}{\rho^2}\gamma_z, \nonumber\\
{\Gamma^2}_{11}&=&\frac{y}{\rho^3}(x^2\gamma_\rho+\rho), \quad
{\Gamma^2}_{12}=\frac{x}{\rho^3}(y^2\gamma_\rho-\rho), \nonumber\\
{\Gamma^2}_{22}&=&\frac{y}{\rho^3}(y^2\gamma_\rho-\rho^2\psi_\rho),
\quad {\Gamma^2}_{23}=\frac{1}{\rho^
2}(y^2\gamma_z-\rho^2\psi_z), \nonumber\\
{\Gamma^2}_{21}&=&\frac{x}{\rho^3}(y^2\gamma_\rho-\rho^2\psi_\rho),
\quad {\Gamma^3}_{33}=\gamma_z-\psi_z, \nonumber\\
{\Gamma^3}_{31}&=&\frac{x}{\rho}(\gamma_\rho-\psi_\rho), \quad
{\Gamma^3}_{32}=\frac{y}{\rho}(\gamma_\rho-\psi_\rho).
\end{eqnarray}
The corresponding non-vanishing components of the torsion tensor
are
\begin{eqnarray}
{T^0}_{01}&=&-\frac{x}{\rho}\psi_\rho=-{T^0}_{10}, \quad
{T^0}_{02}=-\frac{y}{\rho}\psi_\rho=-{T^0}_{20}, \nonumber\\
{T^0}_{03}&=&-\psi_z=-{T^0}_{30}, \quad
{T^1}_{12}=\frac{y}{\rho^2}
(\rho\psi_\rho-1)=-{T^1}_{21}, \nonumber\\
{T^1}_{13}&=&\frac{1}{\rho^2}(\rho^2\psi_z-x^2\gamma_z)=-{T^1}_{31},\quad
{T^2}_{12}=\frac{x}{\rho^2}
(1-\rho\psi_\rho)=-{T^2}_{21}, \nonumber\\
{T^2}_{13}&=&{T^1}_{23}=-\frac{xy}{\rho^2}\gamma_z=-{T^2}_{31}=-{T^1}_{32}, \nonumber\\
{T^3}_{32}&=&\frac{y}{\rho}(\psi_\rho-\gamma_\rho)=-{T^3}_{23},\quad
{T^3}_{31}=\frac{x}{\rho}(\psi_\rho-\gamma_\rho)=-{T^3}_{13}, \nonumber\\
{T^2}_{23}&=&\frac{1}{\rho^2}(\rho^2\psi_z-y^2\gamma_z)=-{T^2}_{32}.
\end{eqnarray}
The required components of the Freud's superpotential are
\begin{eqnarray}
{U_0}^{01}&=&
\frac{x}{2\rho}(2\psi_\rho-\gamma_\rho-\frac{1}{\rho}),
\nonumber\\
{U_0}^{02}&=&\frac{y}{2\rho}(2\psi_\rho-\gamma_\rho-\frac{1}{\rho}),
\nonumber\\
{U_0}^{03}&=&\frac{1}{2}(2\psi_z-\gamma_z).
\end{eqnarray}
When we use these values in Eq.(7), we obtain energy-momentum
density components given in table (1)

\vspace{0.5cm}

{\bf {\small Table 1.}} {\small Energy-Momentum(E-M) densities in
different prescriptions} (i=1,2,3)
\begin{center}
\begin{tabular}{|l|l|l|}
\hline{\bf Prescription}&{\bf E. density}&{\bf M. density}\\
\hline Einstein & $
\begin{array}{c}
hE_0^0=-\frac{1}{8\pi}(\gamma_{\rho\rho}+\gamma_{zz}
+\frac{1}{\rho}\gamma_\rho)
\end{array}
$ & $hE_i^0=0$,\\
\hline Landau-Lifshitz & $
\begin{array}{c}
hL^{00}=\frac{e^{2(\gamma-2\psi)}}{8\pi}
[\frac{4}{\rho}\psi_\rho-\frac{2}{\rho}\gamma_\rho-2(\gamma_\rho-2\psi_\rho)^2\\
-2(\gamma_z-2\psi_z)^2-(
\gamma_{\rho\rho}+\gamma_{zz}+\frac{1}{\rho}\gamma_\rho)]
\end{array}
$ & $hL^{0i}=0$, \\
\hline Bergmann & $
\begin{array}{c}
hB^{00}=\frac{e^{-2\psi}}{8\pi}
[2\{\gamma_\rho\psi_\rho+\gamma_z\psi_z+
\frac{1}{\rho}\psi_\rho\\
~~~~-2(\psi_\rho^2+\psi_z^2)\} -(\gamma_{\rho\rho}+\gamma_{zz}+
\frac{1}{\rho}\gamma_\rho)]
\end{array}
$ & $hB^{i0}=0$, \\
\hline
\end{tabular}
\end{center}
We see that energy takes a well-defined and definite form in each
prescription while momentum becomes constant.

\subsection{Energy-Momentum Densities in M$\ddot{o}$ller Prescription}

By following the same procedure as given in [50,51], we can write
tetrad of the metric (16) as
\begin{equation}
{h^a}_\mu=\left\lbrack\matrix { e^\psi   &&&   0    &&&   0    &&&
0 \cr 0  &&& e^{\gamma-\psi} cos\theta &&& -\rho e^{-\psi}
sin\theta &&& 0 \cr 0 &&& e^{\gamma-\psi} sin\theta &&& \rho
e^{-\psi} sin\theta &&& 0 \cr 0        &&&   0 &&& 0 &&&
e^{\gamma-\psi} \cr } \right\rbrack
\end{equation}
with its inverse
\begin{equation} {h_a}^\mu=\left\lbrack\matrix {
e^{-\psi}   &&   0    &&   0    && 0 \cr 0        &&
e^{\psi-\gamma} cos\theta && -\frac{1}{\rho}e^\psi sin\theta && 0
\cr 0 &&  e^{\psi-\gamma} sin\theta && \frac{1}{\rho}e^\psi
cos\theta && 0\cr 0 &&   0 && 0 && e^{\psi-\gamma} \cr }
\right\rbrack.
\end{equation}
Using Eqs.(26) and (27) in Eq.(2), we get the following
non-vanishing components of the Weitzenb$\ddot{o}$ck connection
\begin{eqnarray}
{\Gamma^0}_{01}&=&\psi_\rho, \quad
{\Gamma^0}_{03}=-{\Gamma^2}_{23}=\psi_z,
\quad\quad {\Gamma^2}_{21}= \frac {1}{\rho}-\psi_\rho , \nonumber\\
{\Gamma^1}_{11}&=&{\Gamma^3}_{31}= \gamma_\rho-\psi_\rho,\quad
{\Gamma^2}_{12}= \frac {1}{\rho}e^\gamma, \quad
{\Gamma^1}_{22} =-\rho e^{-\gamma}, \nonumber\\
{\Gamma^1}_{13}&=&{\Gamma^3}_{33}= \gamma_z- \psi_z.
\end{eqnarray}
The corresponding components of the torsion tensor and the basic
vector field will become
\begin{eqnarray} {T^0}_{01}&=&-\psi_\rho, \quad
{T^0}_{03}={\Gamma^2}_{32}=-\psi_z,
\quad\quad {T^2}_{12}= \frac {1}{\rho}(1-e^\gamma)-\psi_\rho, \nonumber\\
{T^1}_{13}&=&{T^0}_{01}=\psi_z- \gamma_z,\quad {T^3}_{31}=
\psi_\rho- \gamma_\rho
\end{eqnarray}
and
\begin{eqnarray}
\Phi^1=e^{2(\psi-\gamma)}\{\gamma_\rho-\psi_\rho+\frac{1}{\rho}(1-e^\gamma)\},
\quad \Phi^3=e^{2(\psi-\gamma)}(\gamma_z-\psi_z)
\end{eqnarray}
respectively. The required non-vanishing components of the
superpotential in M$\ddot{o}$ller tetrad theory are
\begin{eqnarray}
{U_0}^{01}&=&
\frac{1}{\kappa}(2\rho\psi_\rho-\rho\gamma_\rho+e^\gamma-1),\nonumber\\
{U_0}^{03}&=&\frac{\rho}{\kappa}(2\psi_z-\gamma_z).
\end{eqnarray}
Substituting these results in Eq.(15) and $c,G=1$, it yields
energy and momentum densities in M$\ddot{o}$ller's prescription
\begin{eqnarray}
\Xi_0^0&=&\frac{1}{8\pi}(2\rho {\psi_z}^2+\gamma_\rho
e^\gamma),\nonumber\\ \Xi_i^0&=&0, \quad \Xi_0^i=0.
\end{eqnarray}
This shows that momentum becomes constant and the energy density
turns out as a definite and well-defined quantity. If we take
$\gamma_\rho e^\gamma=-2\rho {\psi_z}^2$ then energy also becomes
constant which coincides with the energy in GR [40-42]. It is
worth mentioning here that the results are independent of the
coupling constant $\lambda$, that is, these results are valid for
any teleparallel theory.

The Curzon metric [52] is a special case of static axially
symmetric spacetimes and can be obtained by substituting
\begin{equation}
\gamma(\rho, z)=-\frac{m^2\rho^2}{2(\rho^2+z^2)^2}, \quad
\psi(\rho,z)=-\frac{m}{\sqrt{\rho^2 + z^2}}.
\end{equation}
in Eq.(16). The energy and  momentum density components turn out
in a simple form as given in the table (2).


{\bf {\small Table 2.}} {\small Energy-Momentum(E-M) densities of
the Curzon Metric} ($i=1,2,3$)

\vspace{0.5cm}

\begin{center}
\begin{tabular}{|l|l|l|}
\hline{\bf Prescription}&{\bf E. density}&{\bf M. density}\\
\hline Einstein & $
\begin{array}{c}
hE_0^0=\frac{m^2z^2}{4\pi r^6}\\
\end{array} $&
$hE^0_i=0$ \\
\hline Landau-Lifshitz & $
\begin{array}{c}
hL^{00}=\frac{me^{2(\gamma-2\psi)}}{4\pi r^3}
[2-\frac{m^3\rho^2}{r^5}-\frac{4m^2\rho^2}{r^4} -\\
~~~~~~~~~~~~~~~~~~~~~~~~~~~~~~ \frac{3m\rho^2}{r^3}-\frac{2m}{r}]
\end{array}
$ & $hL^{0i}=0$ \\
\hline Bergmann & $
\begin{array}{c}
hB^{00}=\frac{m e^{-2\psi}}{4\pi r^3}
[\frac{m^2\rho^2}{r^4}-\frac{m\rho^2}{r^3}-\frac{m}{r}+1]
\end{array}
$ & $hB^{0i}=0 $\\
\hline M$\ddot{o}$ller & $
\begin{array}{c}
\Xi_0^0=\frac{m^2\rho}{8\pi r^6} [e^\gamma(\rho^2-z^2)+2z^2]
\end{array}
$ & $h\Xi^0_i=0 $\\
\hline
\end{tabular}
\end{center}
Here $\gamma(\rho, z)$ and $\psi(\rho, z)$ are given by Eq.(32),
and $r=\sqrt{\rho^2+z^2}$ . This table gives the energy-momentum
distribution for the Curzon metric in four different
prescriptions.

\section{Summary and Discussion}

The problem of localization of energy has been re-considered in
the framework of TPG by many scientists. The authors [30-36]
showed that energy-momentum can also be localized in this theory.
It has been shown that the results of the two theories can agree
with each other. M$\ddot{o}$ller found that a tetrad description
of a gravitational field equation allows a more satisfactory
treatment of the energy-momentum complex than does GR. Vargas [32]
found that the total energy of the closed
Friedmann-Robertson-Walker spacetime is zero by using teleparallel
version of Einstein and Landau-Lifshitz complexes which agreed
with the results of GR [37]. Recently, Sharif and Jamil [38]
evaluated the energy-momentum distribution of Lewis-Papapetrou
spacetime by using M$\ddot{o}$ller's prescription and found that
the results of TPG and GR [39] are not consistent.

In this paper, we have explored the energy-momentum distribution
for static axially symmetric spacetimes by using the TP version of
Einstein, Landau-Lifshitz, Bergmann and M$\ddot{o}$ller's
prescriptions. We see from the table 1 that the energy density
turns out to be different but momentum becomes constant in each
prescription. Further, the expressions for energy density do not
coincide with those given in the framework of GR [40-42] but
momentum is the same. Finally, we have considered the Curzon
metric, a special case of the Weyl metrics. This also leads to
different expressions for the energy density but same for the
momentum. It is interesting to note from table 2 that for the
Curzon metric energy also becomes constant in the limiting case
when $r\rightarrow\infty$ and hence coincides with GR. While the
energy density in each case will diverge at $r=0$, that is, along
$\phi-$axis.

In recent papers [26-28,39-42], Sharif and his collaborators used
different prescriptions to determine the energy-momentum
distribution for various spacetimes in GR. These results do not
coincide for any of the prescriptions. Here we have used the TP
version of different energy-momentum complexes and found that the
energy density is different for the four prescriptions but the
momentum becomes constant. It is mentioned here that these results
turn out to be the same under the limiting case of the Curzon metric
which is a special solution of the Weyl metrics.

We would like to mention here that the results of energy-momentum
distribution for the Weyl metrics are not surprising. This justifies
that different energy-momentum complexes, which are pseudo-tensors,
are not covariant objects. This is in accordance with the
equivalence principle [1] which implies that the gravitational field
cannot be detected at a point. This supports the well-defined
proposal developed by Cooperstock [9] and verified by many authors
[26-28,39-42].

\newpage
\vspace{0.5cm}

{\bf Acknowledgment}

\vspace{0.5cm}

We would like to acknowledge Higher Education Commission Islamabad,
Pakistan for its financial support through the {\it Indigenous PhD
5000 Fellowship Program Batch-I}.

\vspace{0.5cm}

{\bf \large References}

\begin{description}

\item{[1]} Misner, C.W., Thorne, K.S. and Wheeler, J.A.: \textit{Gravitation}
(Freeman, New York, 1973).

\item{[2]} Trautman, A.: \textit{Gravitation}: \textit{An introduction to Current
           Research}, ed. Witten, L. (Wiley, New York, 1962).

\item{[3]} Landau, L.D. and Lifshitz, E.M.: \textit{The Classical Theory
           of Fields} (Addison-Wesley Press, New York, 1962).

\item{[4]} Papapetrou, A.: \textit{Proc. R. Irish Acad. } \textbf{A52}(1948)11.

\item{[5]} Bergman, P.G. and Thompson, R.: Phys. Rev.
           \textbf{89}(1958)400.

\item{[6]} Tolman, R.C.: \textit{Relativity Thermodynamics and
           Cosmology} (Oxford University Press, Oxford, 1934).

\item{[7]} Weinberg, S.: \textit{Gravitation and Cosmology} (Wiley, New
           York, 1972).

\item{[8]} M$\ddot{o}$ller, C.: Ann. Phys. (N.Y.) \textbf{4}(1958)347.

\item{[9]} Cooperstock, F.I. and Sarracino, R.S.: J. Phys. A: Math.
           Gen. \textbf{11}(1978)877.

\item{[10]} Bondi, H.: \textit{ Proc. R. Soc. London } \textbf{A427}(1990)249.

\item{[11]} Penrose, R.: \textit{Proc. Roy. Soc. London }
\textbf{A388}(1982)457.

\item{[12]} Penrose, R.: GR10 Conference eds. Bertotti, B., de Felice, F. and Pascolini,
            A. Padova \textbf{1}(1983)607.

\item{[13]} Brown, J.D. and York Jr., J.W.: Phys. Rev. \textbf{D47}(1993)1407.

\item{[14]} Hayward, S.A.: Phys. Rev. \textbf{D497}(1994)831.

\item{[15]} Bergqvist, G.: Class. Quantum Gravit.  \textbf{9}(1992)1753.

\item{[16]} Chang, C.C., Nester, J.M. and Chen, C.: Phys. Rev. Lett.
            \textbf{83}(1999)1897.

\item{[17]} Xulu, S.S.: Int. J. Mod. Phys. \textbf{A15}(2000)2979.

\item{[18]} Xulu, S.S.: Mod. Phys. Lett. \textbf{A15}(2000)1151.

\item{[19]} Xulu, S.S.: Astrophys. Space Sci. \textbf{283}(2003)23.

\item{[20]} Virbhadra, K.S.: Phys. Rev. \textbf{D42}(1990)2919.

\item{[21]} Virbhadra, K.S. and Parikh, J.C.: Phys. Lett.  \textbf{B317}(1993)312.

\item{[22]} Virbhadra, K.S. and Parikh, J.C.: Phys. Lett.  \textbf{B331}(1994)302.

\item{[23]} Rosen, N. and Virbhadra, K.S.: Gen. Relativ. Gravit.  \textbf{25}(1993)429.

\item{[24]} Aguirregabiria, J.M., Chamorro, A. and Virbhadra, K.S.: Gen. Relativ.
            Gravit. \textbf{28}(1996)1393.

\item{[25]} Virbhadra, K.S.: Phys. Rev. \textbf{D60}(1999)104041.

\item{[26]} Sharif, M.: Int. J. Mod. Phys. \textbf{A17}(2002)1175.

\item{[27]} Sharif, M.: Int. J. Mod. Phys. \textbf{A18}(2003)4361.

\item{[28]} Sharif, M.: Int. J. Mod. Phys. \textbf{D13}(2004)1019.

\item{[29]} Radinschi, I.: Mod. Phys. Lett. \textbf{A16}(2001)673.

\item{[30]} Mikhail, F.I., Wanas, M.I., Hindawi, A. and Lashin, E.I.: Int. J. Theo.
            Phys. \textbf{32}(1993)1627.

\item{[31]} Nashed, G.G.L.: Phys. Rev. \textbf{D66}(2002)060415.

\item{[32]} Vargas, T.: Gen. Relativ. Gravit. \textbf{30}(2004)1255.

\item{[33]} Salti, M., Havare, A.: Int. J. of Mod. Phys. \textbf{A20}(2005)2169.

\item{[34]} Aydogdu, O. and Salti, M.: Astrophys. Space Sci. \textbf{229}(2005)227.

\item{[35]} Aydogdu, O., Salti, M. and Korunur, M.: Acta Phys. Slov. \textbf{55}(2005)537.

\item{[36]} Salti, M.: Astrophys. Space Sci. \textbf{229}(2005)159.

\item{[37]} Rosen, N.: Gen. Relativ. Gravit. \textbf{26}(1994)323.

\item{[38]} Sharif, M. and Amir, M.J.: Mod. Phys. Lett. \textbf{A22} (2007)425.

\item{[39]} Sharif, M. and Azam, M.: \emph{Energy-Momentum Distribution
            of the Wyel-Lewis-Papapetrou and the Levi-Civita Metrics}
            Int. J. Mod. Phys. {\bf A} (to appear 2007).

\item{[40]} Sharif, M. and Fatima, T.: Nouvo Cim. \textbf{B120}(2005)533.

\item{[41]} Sharif, M. and Fatima, T.: Int. J. Mod. Phys. {\bf A20}(2005)4309.

\item{[42]} Sharif, M. and Fatima, T.: Astrophys. Space Sci. {\bf 302}(2006)217.

\item{[43]} Aldrovendi, R. and Pereira, J.G.: {\it An Introduction to
            Gravitation Theory} (preprint).

\item{[44]} Hayashi, K. and Shirafuji, T.: Phys. Rev. {\bf D19}(1979)3524.

\item{[45]} Hehl, F.W., McCrea, J.D., Mielke, E.W. and Neeman, Y.: Phys. Rep.
            {\bf 258}(1995)1.

\item{[46]} Aldrovandi and Pereira, J.G.: {\it An Introduction to
            Geometrical Physics} (World Scientific, 1995).

\item{[47]} Kramer, D., Stephani, H., Herlt, E. and MacCallum, M.:
            \emph{Exact Solutions of Einstein's Field Equations}
            (Cambridge University Press, 1985).

\item{[48]} Synge, J.L.: \emph{Relativity: The General theory}
            (North-Holland Pub. Co. Amsterdam, 1960).

\item{[49]} Hernandez-Pastora1, J. L.  and Martin, J.:  Gen. Relativ. Gravit.
            \textbf{26}(1994)877.

\item{[50]} Pereira, J.G., Vargas, T. and Zhang, C.M.: Class. Quantum Grav.
            {\bf 18}(2001)833.

\item{[51]} Sharif, M. and Amir, M.J.: Gen. Relativ. Gravit. \textbf{38}(2006)1735.

\item{[52]} Curzon, H.E.J.: \textit{Proc. Math. Soc. London} \textbf{23}(1924)477.

\end{description}
\end{document}